\documentclass[12pt]{article}
\usepackage{lmodern}
\usepackage[english]{babel}
\usepackage[T1]{fontenc}
\usepackage[latin9]{inputenc}
\usepackage{geometry}
\usepackage{amsmath}
\usepackage{amsthm}

\newtheorem{lm}{Lemma}
\newtheorem{thm}{Theorem}
\newtheorem{df}{Definition}

\usepackage{amssymb}

\DeclareMathOperator{\const}{const}

\newcommand{\R}{{\mathbb R}}
\newcommand{\C}{{\mathbb C}}
\newcommand{\Z}{{\mathbb Z}}

\newcommand{\D}{{\mathcal D}}
\newcommand{\hilb}{{\mathcal H}}

\newcommand{\s}{{LQC}}
\newcommand{\w}{{WDW}}

\newcommand{\bmat}[1]{\left(\begin{array}{cc} #1 \end{array}\right)}

\usepackage{graphicx}
\usepackage{color}
\usepackage{xcolor}
\usepackage{authblk}
\usepackage{etoolbox}
\usepackage{subfigure}
\usepackage{helvet}
\usepackage{float}
\usepackage[makeroom]{cancel}

\begin{document}

\title{The volume operator in loop quantum cosmology}
\author[1]{Wojciech Kami{\'n}ski}
\affil[1]{\small Instytut Fizyki Teoretycznej, Wydzia{\l} Fizyki, Uniwersytet Warszawski, ul. Pasteura 5 PL-02093 Warszawa, Poland}
\maketitle
\abstract{We show that for basically all states, the evolution of the expectation value of the volume operator in Wheeler-DeWitt and various homogenous isotropic loop quantum comology (LQC) models ($k=0$, coupled to massless scalar field and without cosmological constant) is ill-defined. The expectation value of the volume operator become instantonously infinite during the evolution. The effect is produced by a very long tail in the volume spectrum, that is however semiclassically small and beyond reach of numerical simulations (that is why it went unnoticed so far). This is not necessarily problem of theory but rather it suggests that one should be extremely careful with unbounded observables.}

\section{Introduction}

In this note we will show that one should be careful with computing expectation values of apparently innocent observables in LQC \cite{Bojowald2005, Ashtekar2011, Bojowald2001, Ashtekar2006b, Ashtekar2003, Ashtekar2006, Ashtekar2006a} (as for example the volume operator $V$ that is of utmost importance) because they might be ill-defined in the evolved states. 
The obvious and reasonable remedy is to restrict to bounded observables like some functions of $V$, although in many cases functions like $\ln V$ are good enough. This is already a standard procedure for models with positive cosmological constant \cite{Pawlowski2016}. In this case the volume diverges already at the semiclassical level and it is physically well understood \cite{Pawlowski2012, Kaminski2009b}.

We will deal with various LQC models (including APS \cite{Ashtekar2006b}, sLQC \cite{Ashtekar2007}, MMO \cite{Martin-Benito2009} and sMMO \cite{MenaMarugan2011}) and the WDW model ($k=0$ without cosmological constant). 
There is no such issue for the models with negative cosmological constant, because every eigenstate of the Hamiltonian belongs to the domain of the volume operator. The same is true for APS $k=1$ model \cite{Ashtekar2007a} and we do not expect this phenomena to occure there.
This issue is unrelated to the known non-selfadjointness of the models with positive cosmological constant \cite{Kaminski2009a}. As the recently introduced models \cite{Dapor2018, Yang2009, Garcia2019} exhibits similarity to both these cases \cite{Assanioussi2018}, it is not clear what is the fate of our result in this modified setup.

The problem with finding suitable states in the volume domain was already mentioned in \cite{Martin-Benito2019}. Indeed, for both WDW and these LQC models we will prove that there are basically no states such that expectation value of the volume is finite under the evolution (see section \ref{intro-sec} for notation). 

\begin{thm}\label{thm}
The only vector $\phi\in{\mathcal H}_{\w}$ that satisfies
\begin{equation}
\exists t\not=0 \text{ such that both } \phi,\ e^{it\sqrt{Q}_{\w}}\phi\in D(V_{\w}^{\frac{1}{2}})
\end{equation}
is the zero vector $\phi=0$. Similarly
if vector $\phi\in{\mathcal H}_{\s}$ satisfies
\begin{equation}
\exists t\not=0 \text{ such that both } \phi,\ e^{it\sqrt{Q_{\s}}_+}\phi\in D(V_{\s}^{\frac{1}{2}})
\end{equation}
then $\phi$ is a null eigenvector of $Q_\s$, $Q_\s\phi=0$.
\end{thm}

Let us notice that there are no null eigenvectors for both APS and sLQC thus in this case the only such vector is the zero vector again. As the minimal assumptions on the state for expectation of the volume to be well-defined is exactly $\psi\in D(V^{\frac{1}{2}})$ we see that evolution of the volume is not well-defined (the volume become immediately infinity).

We will show (on example of WDW model) that the reason for this behaviour is some tail in the volume spectrum that is produced during evolution from low energy spectrum of the state. For the Gaussian state peaked at energy $p_0$ it is
\begin{equation}
\rho(v)=\frac{C\sqrt{\sigma}t^2 e^{-2\sigma p_0^2}}{v\ln^4v}+O\left(\frac{1}{v\ln^5v}\right)
\end{equation}
This tail is in some sense small in the classical limit due to the factor $e^{-2\sigma p_0^2}$, but it is also nonintegrable if one computes expectation value of the volume.

The occurence of the term is rooted in the square root in the definition of the Hamiltonian. In some models like sLQC the Hamiltonian can be written as $|A|$ where $A$ is a difference operator. If the initial state is peaked at the positive frequencies solutions then the difference between state evolved with the real Hamiltonian and the Hamiltonian without absolute value is small (it vanishes in the semiclassical limit). 
However as volume operator is unbounded it can still be infinite even on something extremely small. Evolution with the Hamiltonian without absolute value exhibits no problems (that is known in sLQC model).  This is also the reason why we expect that the issue is absent in models with dust time like \cite{Husain2012}, but it might be present in \cite{Pawlowski2014}.

On the other hand, in numerical simulations one always deals with finite precision computations and finite integration domains. In order for the tail to be visible the range of volume integration need to be of order $e^{2\sigma p_0^2}$ that is far beyond the reach for assumed spreads $\sigma$ and mean value $p_0$ of the energy.

\section{Models}\label{intro-sec}

Let us consider a following constraint \cite{Ashtekar2003, Ashtekar2006}
\begin{equation}
p_\phi^2-Q=0,
\end{equation}
where the operator $Q$ commutes with $p_\phi$ and $\phi$ operators (it acts on geometric variable Hilbert space $\hilb^{geom}$). The initial physical Hilbert space $\hilb^{tot}$ consists of two copies of the geometric kinematical Hilbert space projected into nonnegative part of spectrum of $Q$\footnote{The question if one should include null vectors $P_0(Q_\s)$ is disputable.}.
\begin{equation}
\hilb^{tot}=\hilb\oplus \hilb,\quad \hilb=P_{\lambda\geq 0}(Q)(\hilb^{geom}).
\end{equation}
In the models under consideration $Q\geq 0$ thus $\hilb=\hilb^{geom}$.
There are two sectors of the solutions
\begin{equation}
p_\phi=\pm \sqrt{Q}_+,
\end{equation}
where $\sqrt{Q}_+=\sqrt{|Q|}P_{\lambda\geq 0}(Q)$. The relational observables (geometric observables at the given value of $\phi$) preserve this two sectors\footnote{If we obtain observables by group averaging technique then there exist orderings that preserves and ordering that mixes sectors \cite{Kaminski2009b}.}.
In LQC one restricts physical Hilbert space to positive energy solutions and then the physical Hamiltonian is given by $\sqrt{Q}_+$ due to superselection rule \cite{Ashtekar2007}. 

\subsection{Wheeler-DeWitt model}

We consider version of WDW operator on $\R_+$. The standard approach\footnote{We are working in the ''natural'' units (for example $4v_0=1$) and we skip constants. They are unimportant for the issue addressed in this paper and can be easily restored.} is
\begin{equation}
\hilb_\w=L^2\left(\R_+, \frac{dv}{v}\right),\quad Q_\w=-\left(v\partial_v\right)^2.
\end{equation}
The volume operator $V_\w$ is the multiplication operator by $v$.

\subsection{Loop quantum cosmology}

We will now describe standard LQC. Our description (where we skip again constants) covers APS and sLQC models directly (MMO and sMMO need to be rewritten in the $B(v)$ scalar product). We consider here only sector $\epsilon=0$ with parity assumption, but we expect that our method can be extend beyond that. The Hilbert space is\footnote{Inclusion of $v=0$ does not alter our result.}
\begin{equation}
\hilb_\s=\{\phi\colon \Z_+\rightarrow \C,\ \|\phi\|<\infty\},\quad \|\phi\|^2=\sum_{v\in \Z_+} B(v)|\phi(v)|^2
\end{equation}
and the operator $Q_\s$ is defined by
\begin{equation}
Q_\s\phi(v)=-B(v)^{-1}\left(C^+(v)\phi(v+1)+C^0(v)\phi(v)+C^-(v)\phi(v-1)\right),
\end{equation}
where we assumed $\phi(v)=0$ for $v\leq 0$. The functions $B$, $C^\pm$ and $C^0$ admits an expansion in inverse powers of $v$ and they are assumed to satisfy
\begin{align}
B(v)&=\frac{1}{v}+O\left(\frac{1}{v^2}\right),\\
C^\pm(v)&=\alpha \left(v\pm \frac{1}{2}\right)+\beta+\frac{k^\pm}{v}+O\left(\frac{1}{v^2}\right),\\
C^0(v)&=-2\alpha v-2\beta+\frac{k^0}{v}+O\left(\frac{1}{v^2}\right),
\end{align}
where we assume:
\begin{enumerate}
\item $C^-(v)=C^+(v-1)\in\R$ and $C^0(v)\in\R$ and that $C^\pm(v)\not=0$,
\item $B(v)\not=0$,
\item $\alpha>0$ and $\beta\in\R$,
\item $k^++k^-+k^0=0$ (without this assumption the eigenfunction approaches WDW eigenfunctions with modified energy).
\end{enumerate}
The operators that can be written for $v$ positive in the form
\begin{equation}
B(v)^{-1}(h-h^*)A(v)(h-h^*),
\end{equation}
where $h$ is a shift by $\frac{1}{2}$ and
\begin{equation}
A(v)=\alpha v+\beta -\frac{k^0}{2v}+O\left(\frac{1}{v^2}\right)
\end{equation}
satisfy the assumptions. This is the case of both APS and sLQC. Our assumptions are however satisfied also by MMO and sMMO prescriptions  (see \cite{MenaMarugan2011}\footnote{Operator $\Theta$ in \cite{MenaMarugan2011} differs from $Q_\s$ due to another scalar product.}).

The volume operator $V_\s$ is multiplication operator by $v$. 

\section{Domain of the volume operator}

We will base proof of the theorem on the following observation. Let us define

\begin{df}
Let $\D_\gamma$ denote the space of holomorphic functions
\begin{equation}
f\colon \{z\in \C\colon \Im z\in(-\gamma,0)\}\rightarrow \C,
\end{equation}
such that there exists a measureable function $\tilde{f}\colon \R\rightarrow \C$ that for any $g\in C^\infty_0(\R\setminus \{0\})$
\begin{equation}
\lim_{\epsilon\rightarrow 0_+} \int dp\ g(p)f(p-i\epsilon)=\int dp\ g(p)\tilde{f}(p).
\end{equation}
\end{df}

In fact the last equality defines $\tilde{f}(p)$ as a distributional boundary value limit for $p\not=0$. 

\begin{thm}\label{thm-ex}
Let us assume that there exist two functions
\begin{equation}
f_0, f_t\in \D_\epsilon
\end{equation}
such that (almost everywhere on $\R\setminus\{0\}$) the boundary values satisfy
\begin{equation}
\tilde{f}_t(p)=e^{it|p|}\tilde{f}_0(p),
\end{equation}
then $f_t=f_0=0$.
\end{thm}

The proof is based on the following

\begin{lm}\label{lm-zero}
Let $\tilde{f}$ be a distributional boundary limit of a holomorphic function $f\in \D_\epsilon$. If on some interval $I\subset \R\setminus \{0\}$, $\tilde{f}=0$ then $f=0$ everywhere.
\end{lm}

\begin{proof}
Let us consider an extension of the function to the positive imaginary strip by
\begin{equation}
f(p+iq)=\overline{f(p-iq)}
\end{equation}
There exists also a distributional boundary value from the positive imaginary side. Moreover on $I$ both boundary values are equal. By the edge of the wedge theorem \cite{Rudin-edge} the function extends analytically through $I$. The extended function is zero because it is zero on an interval.
\end{proof}

We can now prove theorem:

\begin{proof}
Suppose that $f_0$ is nontrivial.
We know that boundary value function $\tilde{f}_0$ is not identically zero both for $p>0$ and $p<0$ (see lemma \ref{lm-zero}). Let us now consider a holomorphic (in the strip) functions
\begin{equation}
g_t(z)=f_t(z)-e^{izt}f_0(z).
\end{equation}
For $p>0$ its boundary value is $\tilde{f}_t(p)-e^{ipt}\tilde{f}_0(p)=0$ thus $g_t=0$. Let us now compute its boundary value for $p<0$
\begin{equation}
\tilde{f}_t(p)-e^{ipt}\tilde{f}_0(p)=\left(e^{-ipt}-e^{ipt}\right)\tilde{f}_0(p).
\end{equation}
However it needs to be zero almost everywhere. We obtain $\tilde{f}_0(p)=0$ for $p<0$ as $t\not=0$. This contradicts what we stated at the beginning and thus $f_0=0$. Similarly $f_t=0$ as the boundary value vanishes.
\end{proof}

\subsection{Volume operator in WDW theory}

We can rescale the functions by $v^{-\frac{1}{2}}$ and then
\begin{equation}
Q_\w=\left(\frac{i}{2}(v\partial_v+\partial_v v)\right)^2,\quad \hilb=L^2(\R_+, dv)
\end{equation}
and the operator $V_\w=v$.

We now consider evolution given by $H_\w=\sqrt{Q_\w}$.
Let us change variables $x=\ln v$ and then (using again rescaling of the function to obtain standard scalar product)
\begin{equation}
H_\w=|p|=|i\partial_x|,\quad V_\w=e^x.
\end{equation}
The domain of the self-adjoint operator $V_\w^{\frac{1}{2}}$ are all $L^2(\R)$ functions $f$ that satisfy
\begin{equation}
\int_{-\infty}^\infty dx\ e^{x} |f|^2<\infty.
\end{equation}

Let us state some property:

\begin{lm}\label{lm:dom}
Let $\phi(x)\in D(e^{\gamma x})$ for some $\gamma>0$ then $\hat{\phi}(p)\in L^2(\R)$ is a boundary value (in distributional sense) of a holomorphic function defined in $\{z\colon \Im z\in(-\gamma,0)\}$.
\end{lm}

\begin{proof}
Let us notice that the functions
\begin{equation}
\xi_z(x)=\frac{1}{\sqrt{2\pi}}\frac{e^{izx}}{e^{\gamma x}+1}
\end{equation}
belong to $L^2(\R)$ for $\Im z\in (-\gamma,0)$ and they depend holomorphically on $z$. 
We define (as $(e^{\gamma x}+1)\phi\in L^2(\R)$)
\begin{equation}
f(z)=\langle \xi_{-\overline{z}}, (e^{\gamma x}+1)\phi\rangle,
\end{equation}
thus $f(z)$ is analytic for $\Im z\in(-\gamma,0)$. 

Let us notice that for any $g\in C^\infty_0(\R)$ we have (in $L^2$ norm by the dominant convergence theorem)
\begin{equation}
\lim_{\epsilon\rightarrow 0_+} \int dp\ \overline{g(p)}\xi_{-p-i\epsilon}(x)=\frac{1}{e^{\gamma x}+1}\tilde{g}(x),
\end{equation}
where $\tilde{g}(x)$ is the inverse Fourier transform of $\overline{g}$ thus 
\begin{equation}
\lim_{\epsilon\rightarrow 0_+}\int dp\ g(p)f(p-i\epsilon)=\langle \tilde{g},\phi\rangle=\langle \bar{g},\hat{\phi}\rangle=\int dp\ g(p)\hat{\phi}(p).
\end{equation}
Thus the boundary value of $f$ is $\hat{\phi}$.
\end{proof}

Let us notice that lemma \ref{lm:dom} shows that if $\phi$ belongs to $D(V_\w^{\frac{1}{2}})$ then $\hat{\phi}\in \D_{1/2}$. By theorem \ref{thm-ex} there exists no state in the domain of $D(V_\w^{\frac{1}{2}})$ that stays in the domain under the evolution of WDW theory. 

\subsection{Volume operator in LQC}

Let us introduce a notation

\begin{df}
We write $f_p(v)=O_p(v^{\gamma(p)})$ if there exists a continuous $C(p)>0$ for $p\in \left\{z\in \C\colon z\not=0,\ |\Im z|<\frac{1}{2}\sqrt{\alpha}\right\}$ such that
\begin{equation}
|f_p(v)|\leq C(p) v^{\gamma(p)},\quad v\in \Z_+.
\end{equation}
\end{df}
Let us denote a solution to the difference equation (for all $v>0$)
\begin{equation}
(Q_\s-p^2)\phi_p=0,\quad \phi_p(1)=1,
\end{equation}
where $\phi_p\colon \Z_+\rightarrow \C$.

We will now state the important properties of this functions that we will prove later
\begin{enumerate}
\item $\phi_p(v)$ is analytic in $p$ for every $v\geq 1$.
\item $\phi_p(v)=O_p\left(v^{\frac{|\Im p|}{\sqrt{\alpha}}}\right)$,
\item For any $g\in C^\infty_0(\R\setminus \{0\})$ we define
\begin{equation}\label{eq:hilb}
\phi(v)=\int_\R dp g(p)\phi_p(v),
\end{equation}
then $\phi\in \hilb_\s$ and for any $t\in \R$
\begin{equation}\label{eq:u}
e^{it\sqrt{Q_\s}_+}\int_\R dp g(p)\phi_p=\int_\R dp e^{it|p|}g(p)\phi_p.
\end{equation}
\item Vectors of the form 
\begin{equation}
\int_\R dp g(p)\phi_p\in P_{\lambda>0}(Q_\s)\hilb_\s
\end{equation}
for $g\in C^\infty_0(\R\setminus \{0\})$ and they are dense in $P_{\lambda>0}(Q_\s)\hilb_\s$.
\end{enumerate}

The last two properties tell us also that $\phi_p$ for $p\in\R_+$ is the eigenfunction expansion and the positive part of the spectrum is absolutely continous and it is the whole $\R_+$.

Let us now take $\psi\in D(V_\s^\frac{1}{2})$. Let us notice that
\begin{equation}
f_\psi(p)=\sum_{v\in\Z_+} B(v)\phi_p(v)\overline{\psi(v)}
\end{equation}
is locally uniformly absolutely summable for any $p\in \left\{z\in \C\colon z\not=0,\ |\Im z|<\frac{1}{2}\sqrt{\alpha}\right\}$ because $v^{-\frac{1}{2}}\phi_p(v)\in \hilb_\s$ and thus
\begin{equation}
\sum_{v\in\Z_+} B(v)\phi_p(v)\overline{\psi(v)}=
\left\langle V_\s^{\frac{1}{2}}\psi, V_\s^{-\frac{1}{2}}\phi_p\right\rangle
\end{equation}
and it is holomorphic in this domain.

Let us now suppose that
\begin{equation}
\psi_t=e^{-it\sqrt{Q_\s}_+}\psi\in D(V_\s^\frac{1}{2}),
\end{equation}
thus $f_{\psi_t}\in \D_{\frac{1}{2}\sqrt{\alpha}}$ and moreover these functions just extends analytically through the real axis. 

From the properties stated above and absolute summability we see that for any $g\in C^\infty_0(\R\setminus \{0\})$ we have
\begin{align}
&\int_\R dp\ g(p)f_{\psi_t}(p)=\left\langle \psi_t,\int_\R dp\ g(p)\phi_p\right\rangle=\left\langle e^{-it\sqrt{Q_\s}_+} \psi,\int_\R dp\ g(p)\phi_p\right\rangle=\\
&=\left\langle \psi,e^{it\sqrt{Q_\s}_+} \int_\R dp\ g(p)\phi_p\right\rangle=\int_\R dp\ g(p)e^{it|p|}f_{\psi_t}(p).
\end{align}
Thus as distributional limit is equal to pointwise limit
\begin{equation}
f_{\psi_t}(p)=f_{\psi}(p)e^{it|p|}
\end{equation}
for $p\in\R\setminus\{0\}$. From theorem \ref{thm-ex} we obtain $f_{\psi}=0$ and thus $\psi$ is orthogonal to $P_{\lambda>0}(Q_\s)\hilb_\s$ and the only such possible vector in the physical Hilbert space is a null vector of $Q_\s$.


\subsection{Derivation of the properties of $\phi_p$}

In this section we will show the desired properties of functions $\phi_p$. We will use method of the transfer matrix (see \cite{Teschl} and for earlier application to LQC \cite{Kaminski2010a}). 

\begin{lm}
There exists two functions $\psi_p^\pm(v)$ for $p\in\{z\in \C\colon z\not=0,\ |\Im z|<\frac{1}{2}\sqrt{\alpha}\}$ such that
\begin{enumerate}
\item They satisfy 
\begin{equation}
Q_\s\psi_p^\pm(v)=p^2\psi_p^\pm(v),\quad v\not=1.
\end{equation}
\item We have
\begin{equation}\label{eq:asym}
\psi_p^\pm(v)=v^{\pm i \frac{p}{\sqrt{\alpha}}}\left(1+r_\pm^p(v)\right),
\end{equation}
where $r_\pm^p(v)=O_p\left(\frac{1}{v}\right)$ and $r_\pm^p(v+1)-r_\pm^p(v)=O_p\left(\frac{1}{v^2}\right)$.
\item For every $v$, $\psi_p^\pm(v)$ is a holomorphic function.
\end{enumerate}
\end{lm}

\begin{proof}
Let us introduce ($N$ will be specified later)
\begin{equation}
d_p^N(v)=1+\sum_{n=1}^N \frac{b_n(p)}{v^n}.
\end{equation}
We will determine coefficients $b_n$ such that (it will be also $O_p\left(v^{-N-1}\right)$)
\begin{equation}\label{eq:vv}
C^+(v)d_p^N(v+1)d_p^N(v)+(B(v)p^2+C^0(v)) d_p^N(v)+C^-(v)=O\left(\frac{1}{v^{N+1}}\right).
\end{equation}
Let us notice that the terms with $v$ and $1$ in the expansion in $v^{-1}$ vanish identically.  The coefficient at the term with $v^{-1}$ of the equation \eqref{eq:vv} is equal to
\begin{equation}
\alpha b_1^2+k^++k^0+k^-+p^2=0,
\end{equation}
thus with the assumption $k^++k^0+k^-=0$ we get two solutions
\begin{equation}
b_1^\pm=\pm i\frac{p}{\sqrt{\alpha}}.
\end{equation}
The term with $v^{-n}$ has the form
\begin{equation}
\alpha\left(1-n+2b_1^\pm\right)b_n^\pm+ F_n(\{b_k^\pm\colon k<n\})=0.
\end{equation}
where $F_n$ is a polynomial. As $1-n\pm i\frac{2p}{\sqrt{\alpha}}\not=0$ for $|\Im p|<\frac{1}{2}\sqrt{\alpha}$ we can determine the coefficients recursively.

Let us introduce a transfer matrix
\begin{equation}
C_v^p=\bmat{-\frac{B(v)p^2+C^0(v)}{C^+(v)} & -\frac{C^-(v)}{C^+(v)}\\ 1 & 0}
\end{equation}
and auxiliary functions
\begin{equation}
\tilde{\phi}_p^{N\pm}(v)=\prod_{w=w_0}^v d_{p}^{N\pm}(w),\quad 
V_v^p=\bmat{\tilde{\phi}_p^{N+}(v) &\tilde{\phi}_p^{N-}(v)\\ \tilde{\phi}_p^{N+}(v-1) &\tilde{\phi}_p^{N-}(v-1)},
\end{equation}
where $w_0$ is chosen such that the $d_{p}^{N\pm}(w)\not=0$ and $d_{p}^{N+}(w)\not=d_{p}^{N-}(w)$ for $w\geq w_0$ (such $w_0$ is chosen locally in $p$).
Let us notice that
\begin{equation}
V_v^p=\underbrace{\bmat{d_{p}^{N,+}(v)& d_{p}^{N-}(v)\\ 1 & 1}}_{=M_p^N(v)}\bmat{\tilde{\phi}_p^{N+}(v-1) &0\\ 0 &\tilde{\phi}_p^{N-}(v-1)}
\end{equation}
and the simple computation shows that ($\det M_p^N(v)=2i\frac{p}{\sqrt{\alpha}v}+O_p(v^{-2})$)
\begin{equation}
\|M_p^N(v)^{-1}\||=O_p(v).
\end{equation}
Let us now notice that
\begin{equation}
\ln \tilde{\phi}_p^{N\pm}(v)=\sum_{w=w_0}^v \ln d_{N,p}^\pm(w)=\sum_{w=w_0}^v \pm i\frac{p}{\sqrt{\alpha}v}+ O_p\left(\frac{1}{v^2}\right)=\pm i \frac{p}{\sqrt{\alpha}}\ln v+\const+ O_p(v^{-1}),
\end{equation}
thus $|\tilde{\phi}_p^\pm(v)|=O_p(v^{\frac{|\Im p|}{\sqrt{\alpha}}})$ and 
$\tilde{\phi}_p^\pm(v)=Cv^{\pm i \frac{p}{\sqrt{\alpha}}}\left(1+\tilde{r}^\pm(v)\right)$ and $C$ depends analytically on $p$. The error terms satisfy $\tilde{r}^\pm(v)=O_p(v^{-1})$ and $\tilde{r}^\pm(v+1)-\tilde{r}^\pm(v)=O_p(v^{-2})$.

This allows us to estimate
\begin{equation}
\|(V_v^p)^{-1}\|=O_p(v^{1+\alpha^{-\frac{1}{2}}|\Im p|}),\quad \|V_v^p\|=O_p(v^{\alpha^{-\frac{1}{2}}|\Im p|}).
\end{equation}
Moreover from property of $d_{N,p}^\pm$
\begin{equation}
\|C_v^pV_v^p-V_{v+1}^p\|\leq O_p(v^{-N-2})\|V_{v-1}^p\|=O_p(v^{-N-2+\alpha^{-\frac{1}{2}}|\Im p|})
\end{equation}
and finally
\begin{equation}
\|(V_{v+1}^p)^{-1}C_v^pV_v^p-{\mathbb I}\|= O_p(v^{-N-1+2\alpha^{-\frac{1}{2}}|\Im p|}).
\end{equation}
For sufficiently large $N$ it is summable and that means that there exists an invertible limit (it depends analytically on $p$)
\begin{equation}
M_p=\prod_{w=w_0}^\infty (V_{w+1}^p)^{-1}C_w^pV_w^p=\lim_{v\rightarrow \infty} (V_{v+1}^p)^{-1}\left( \prod _{w=w_0}^\infty C_w^p\right)\ V_{w_0}^p
\end{equation}
and it is fastly convergent (like $O_p(v^{-N+2\alpha^{-\frac{1}{2}}|\Im p|})$). Let us notice that this means
\begin{equation}
\prod _{w=w_0}^v C_w^pV_{w_0}^{p}M_p^{-1}=V_{v+1}^p\left({\mathbb I}+O_p(v^{-N+2\alpha^{-\frac{1}{2}}|\Im p|})\right)=
V_{v+1}^p+O_p(v^{-N+3\alpha^{-\frac{1}{2}}|\Im p|}).
\end{equation}
The first row of the lefthand side of the equation are solutions (for $v\geq w_0$) from \eqref{eq:asym} (after normalization). The properties of $r_\pm^p(v)$ follows from expansion of $\tilde{\phi}_p^\pm(v)$. The solutions depends holomorphically on $p$ and moreover as they are uniquely determined by their asymptotic behaviour (error term in asymptotic expansion of the bigger solution is smaller then the second solution).
\end{proof}

From the definition of $\phi_p$ we know that $\phi_p(v)$ is a polynomial in $p^2$ thus it is analytic.
Let us now notice that we can write
\begin{equation}
\phi_p=\alpha^+(p)\psi^+_p+\alpha^-(p)\psi^-_p,
\end{equation}
where $\alpha^\pm$ are holomorphic. As $|\psi^\pm_p|= O_p\left(v^{\pm\alpha^{-\frac{1}{2}}\Im p}\right)$ we have also $\phi_p=O_p(v^{\alpha^{-\frac{1}{2}}|\Im p|})$.

Let us introduce wronskian for two solutions $\phi$, $\psi$
\begin{equation}
w(\phi,\psi)=C^+(v)(\phi(v+1)\psi(v)-\phi(v)\psi(v+1)).
\end{equation}
It is independent of $v$. We can compute 
\begin{align}
&w(\psi^+_p,\psi^-_p)=\lim_{v\rightarrow \infty} C^+(v)(\psi_p^+(v+1)\psi_p^-(v)-\psi_p^+(v)\psi_p^-(v+1))=\\
&=\lim_{v\rightarrow \infty} \alpha v((1+v^{-1})^{ip}(1+r_+^p(v+1))(1+r_-^p(v))-\\
&-(1+v^{-1})^{-ip}(1+r_+^p(v))(1+r_-^p(v+1)))
=2ip\alpha.
\end{align}
The terms with  $r_\pm^p$ cancel.

Let us denote the resolvent kernel for $p^2\notin \R$
\begin{equation}
K_p(w,v)=\left(\frac{1}{Q_\s-p^2}\delta_v\right)(w),
\end{equation}
where $\delta_v$ is the basis vector.

The standard formula for the kernel of the resolvent \cite{Teschl} is given by (for $\pm \Im p>0$) 
\begin{equation}
K_{p}^\pm(w,v)=\frac{1}{w(\psi^\mp_p,\phi_p)}\left\{\begin{array}{ll}
     \psi^\mp_p(w)\phi_p(v) & v<w\\
     \phi_p(w)\psi_p^\mp(v) & v\geq w
     \end{array}\right..
\end{equation}
For any $\tilde{g}\in C_0^\infty(\R_+)$ we have
\begin{equation}
\tilde{g}(Q_\s)\delta_v=
\lim_{\epsilon\rightarrow 0_+} \int_0^\infty  dx\ \tilde{g}(x)\frac{1}{2\pi i}\left(\frac{1}{Q_\s-x-i\epsilon}-\frac{1}{Q_\s-x+i\epsilon}\right)\delta_v.
\end{equation}
Taking the pointwise limit we get
\begin{equation}
\left(\tilde{g}(Q_\s)\delta_v\right)(w)=
\int_0^\infty  dx\ \tilde{g}(x)\frac{1}{2\pi i}\left(K_{\sqrt{x}}^+(w,v)-K_{\sqrt{x}}^-(w,v)\right).
\end{equation}
Let us notice that for $p\in\R$ from reality of $\phi_p$ and large $v$ behaviour of $\psi^\pm_p$ we have $\alpha^+(p)=\overline{\alpha^-(p)}$  and thus they are nonzero for $p\not=0$. For $p\not=0$
\begin{equation}
w(\psi^\mp_p,\phi_p)=\mp \alpha_p^\pm 2ip\alpha.
\end{equation}
We get
\begin{equation}
K_{p}^+(w,v)-K_{p}^-(w,v)=\frac{i}{2p\alpha|\alpha^+_p|^2}\phi_p(v)\phi_p(w)
\end{equation}
and thus
\begin{equation}
\left(\tilde{g}(Q_\s)\delta_v\right)(w)=\int_0^\infty  dp\ \tilde{g}(p^2)\frac{1}{2\pi\alpha|\alpha^+_p|^2}\phi_p(v)\phi_p(w).
\end{equation}
Let us now take a function (let us notice that $\phi_p(1)=1$)
\begin{equation}
\tilde{g}(x)=2\pi\alpha|\alpha^+_{\sqrt{x}}|^2(g(\sqrt{x})+g(-\sqrt{x})),
\end{equation}
then
\begin{equation}
\int_{-\infty}^\infty dp g(p)\phi_p=\tilde{g}(Q_\s)\delta_1\in\hilb_\s.
\end{equation}
Moreover as 
\begin{equation}
e^{it\sqrt{Q_\s}_+}\tilde{g}(Q_\s)\delta_1=\left(e^{it\sqrt{|x|}}\tilde{g}\right)(Q_\s)\delta_1,
\end{equation}
we have also property \eqref{eq:u}.

Vectors of the form $\tilde{g}(Q_\s)\delta_v$ where $\tilde{g}\in C_0^\infty(\R_+)$ are dense in $P_{\lambda>0}(Q_\s)\hilb_\s$. They can be written in the form \eqref{eq:hilb}. This shows the last property.

\section{Example in WDW theory}

We will now explain this puzzling behaviour on the example of the Gaussian state in WDW theory. Let us consider a Gaussian state in momentum representation (as it is usually done in LQC as the momentum eigestates are also eigenfunctions of the Hamiltonian)
\begin{equation}
\hat{\psi}(p)=Ce^{-\sigma (p-p_0)^2+ix_0p},
\end{equation}
where $C=\left(\frac{2\sigma}{\pi}\right)^{\frac{1}{4}}$ is a normalization constant and we assume $p_0>0$.
Let us notice that $\psi\in D(V_\w)$ as it is also a Gaussian. Let us now consider evolved state
\begin{equation}
\hat{\psi}_t(p)=Ce^{-\sigma (p-p_0)^2+i(x_0p+t|p|)}.
\end{equation}
We introduce also a state evolved by a hamiltonian ``without absolute value''
\begin{equation}
\hat{\chi}_t(p)=Ce^{-\sigma (p-p_0)^2+i(x_0p+tp)}.
\end{equation}
Its Fourier transform is a gaussian too.
Let us now notice that
\begin{equation}
\hat{\psi}_t(p)-\hat{\chi}_t(p)=-2i \sin (tp)Ce^{-\sigma (p-p_0)^2+ix_0p}\Theta(-p).
\end{equation}
We have the following fact

\begin{lm}
Let $f(p)$ be a function on the real line satisfying  for $N\in {\mathbb Z}_+$ 
\begin{enumerate}
\item For all $0\leq n\leq N$
\begin{equation}
f^{(n)}(p)\in L^1(\R)\text{ and } C_n=\int_{-\infty}^\infty |f^{(n)}(p)|dp,\quad \lim_{p\rightarrow\pm\infty} f^{(n)}(p)=0,
\end{equation}
\item $f$ is smooth everywhere except $0$ and the limits
\begin{equation}
d_{n,\pm}=\lim_{x\rightarrow 0_\pm}f^{(n)}(p)
\end{equation}
exist for $0\leq n\leq N$,
\end{enumerate}
then the Fourier transform satisfies
\begin{equation}
\hat{f}(x)=\frac{1}{\sqrt{2\pi}}\sum_{n=0}^{N-1} \frac{i^{n-1}\Delta_n}{x^{n+1}}+r_{N}(x),\quad \Delta_{n}=d_{n,+}-d_{n,-},
\end{equation}
where $r_N(x)=o(x^{-N})$ (behaviour at $\pm\infty$) and $|r_N(x)|\leq \frac{C_N}{\sqrt{2\pi}} x^{-N}$. 
\end{lm}

\begin{proof}
Let us consider functions ($|f^{(n)}(p)|$ are integrable)
\begin{equation}
F_\pm^n(x)=\frac{i^n}{x^n}\int_0^\infty dp\ f^{(n)}(\pm p)e^{\pm ipx}.
\end{equation}
Let us notice that $|F_+^n(x)|+|F_-^n(x)|\leq C_n x^{-n}$. By Lebegue Riemann lemma the integrals are vanishing for large $x$ thus also
\begin{equation}
F_\pm^n(x)=o(x^{-n}).
\end{equation}
We can integrate by parts to get 
\begin{equation}
F_\pm^n(x)=\pm\frac{i^{n+1}}{x^{n+1}}d_{n,\pm}+F_\pm^{n+1}(x),
\end{equation}
and summing
\begin{equation}
F_\pm^0(x)=\sum_{n=0}^{N-1} \frac{i^{n+1}}{x^{n+1}}d_{n,\pm}+F_\pm^{N}(x).
\end{equation}
Finally
\begin{equation}
\hat{f}(x)=\frac{1}{\sqrt{2\pi}}( F_+^0(x)+F_-^0(x))=\frac{1}{\sqrt{2\pi}}
\sum_{n=0}^{N-1} \frac{i^{n+1}(d_{n,+}-d_{n,-})}{x^{n+1}}+r_{N}(x),
\end{equation}
where $r_{N}(x)=o(x^{-N})$ and $|r_N(x)|\leq \frac{C_N}{\sqrt{2\pi}} x^{-N}$.
\end{proof}

The function $\hat{\psi}_t-\hat{\chi}_t$ satisfies the assumptions of the lemma with 
\begin{equation}\label{eq-Delta}
\Delta_0=0,\quad \Delta_1 =2itCe^{-\sigma p_0^2+ip_0x_0},\quad
\Delta_2=
O(p_0e^{-\sigma p_0^2})
\end{equation}
and as $\int_x^\infty dx\ x^{n}e^{-x^2}=O(x^{n-1}e^{-x^2})$ we have 
\begin{equation}
C_2=O\left(p_0e^{-\sigma p_0^2}\right),
\end{equation}
thus
\begin{equation}
\psi_t-\chi_t=\frac{1}{\sqrt{2\pi}}\frac{-\Delta_1}{x^2}+r(x),
\end{equation}
where $r(x)=\frac{1}{\sqrt{2\pi}}\frac{-i\Delta_2}{x^3}+o(x^{-3})=O(x^{-3})$ and moreover it is of order $p_0e^{-\sigma p_0^2}$. 
The function
\begin{equation}
e^{x}\left|\frac{\Delta_1}{x^2}+O(x^{-3})\right|^2
\end{equation}
is not integrable at $+\infty$ thus as $\chi_t$ is a gaussian we obtain $\psi_t\notin D(V_\w^{\frac{1}{2}})$.

\subsection{Spectral density of the volume in WDW model}

Let us now analyze $\psi_t(v)$ where $v=e^x$. Let us rewrite an asymptotic expansion in terms of $v$ variable (using change of measure $v^{-\frac{1}{2}}$)
\begin{equation}
\psi_t(v)=\frac{\Delta_1}{\sqrt{2\pi}\sqrt{v}\ln^2v}+O\left(\frac{1}{\sqrt{v}\ln^3v}\right)
\end{equation}
The density $\rho(v)=|\psi_t(v)|^2$ is
\begin{equation}
\rho(v)=\frac{|\Delta_1|^2}{2\pi v\ln^4v}+O\left(\frac{1}{v\ln^5v}\right)
\end{equation}
Let us notice that 
\begin{equation}
|\Delta_1|^2=4t^2\sqrt{\frac{2\sigma}{\pi}} e^{-2\sigma p_0^2}
\end{equation}
and it is small for semiclassically peaked states. The error term consists of the part that is slowly decaying with $v$ (it is nonintegrable if multiplied by $v$), but it is semiclassically small (it is of order $p_0^2e^{-2\sigma p_0^2}$) and the part that is not semiclassically small but it is fastly decaying with $v$. 
Thus although the state might be still peaked at semiclassical values (if we choose right dispersion and take the suitable limit) we see that there is a tail
that is a source of the problem. As long as we take a function of $V$ that is integrable with this tail (for example any bounded function or $\ln V$ or $\ln^2 V$) we can regain semiclassical behaviour in the limit.

\section{Summary}

We showed that the expectation value of the volume $\langle V_\s\rangle_t$ is not well defined in standard LQC theory ($k=0$ and $\Lambda=0$ coupled to the massless scalar field). It is extremely nonclassical behavior, however problems can be avoided if one restricts to the bounded operators like $\operatorname{arctg} V$, that is already a standard practise for LQC with positive comological constant \cite{Pawlowski2016} (we conjecture that in the case of $\Lambda=0$ the expectation value of $\ln V$ is already well behaved). The reason for this puzzling behaviour (as shown on example of WDW model) is the contribution from low energy part of the spectrum that produces a nonintegrable tail. This long tail is however extremely small (beyond reach of numerical simulations and so it went unnoticed). Moreover, it disappears in the semiclassical limit.  We do not know to what extent our result apply also to the models proposed by \cite{Dapor2018, Yang2009}.
Let us also mention that our example shows that the restrictions derived using moments approach \cite{Bojowald2007, Bojowald2012} can be avoided in some models.

Quantum cosmology allows us to gain insight into quantum gravity. With this point of view we would like to propose a solution to the aforementioned issue that is applicable in general loop quantum gravity \cite{Ashtekar-Lewandowski, thiemann}. Considering only bounded operators is the most obvious one. If one nevertheless wants to consider unbounded operators one can argue that avoidance of nonanalytic operations (like a square root) would also lead to better behaved theory. However, there are many nonanalytic operations  in the current constructions of the LQG hamiltonians as well as the volume operator.

On the other hand within the framework of group averaging (with ordering prescription of geometric observables mixing sectors) there is no reason to restrict to the sector of the positive energy solutions. Our result may also indicates that such restriction is not justified. We can conjecture that removing this restriction will solve the issue.

\vskip 0.2cm
\noindent\textbf{Acknowledgements:} We thank Tomasz Paw{\l}owski and Jerzy Lewandowski for useful discussions.

\bibliography{LQC}{}
\bibliographystyle{ieeetr}

\end{document}